\documentstyle[prl,aps,preprint,eqsecnum,rotating,epsfig]{revtex}

\tightenlines

\newcommand{\be}{\begin{equation}}
\newcommand{\ee}{\end{equation}}
\newcommand{\ba}{\begin{eqnarray}}
\newcommand{\ea}{\end{eqnarray}}

\begin{document}
\draft

\title{Cardy-Verlinde Formula and asymptotically \\
flat rotating Charged black holes}
  \author{Jiliang Jing$^*$ \footnotetext[1]
 {email: jljing@hunnu.edu.cn}}
\address{  Institute of Physics and  Department of Physics,
Hunan Normal University,\\ Changsha, Hunan 410081, P. R. China
 \\ and \\ Department of Astronomy and Applied Physics, University of
Science and Technology of China, \\ Hefei, Anhui 230026, P. R.
China }

\maketitle
\begin{abstract}

The Cardy-Verlinde formula is generalized to the asymptotically
flat rotating charged black holes in the Einstein-Maxwell theory
and low-energy effective field theory describing string by using
some typical spacetimes, such as the Kerr-Newman,
Einstein-Maxwell-dilaton-axion, Kaluza-Klein, and Sen black holes.
For the Kerr-Newman black hole, the definition of the Casimir
energy takes the same form as that of the Kerr-Newman-AdS$_4$ and
Kerr-Newman-dS$_4$ black holes, while the Cardy-Verlinde formula
possesses different from since the Casimir energy does not appear
in the extensive energy. The Einstein-Maxwell-dilaton-axion,
Kaluza-Klein, and Sen black holes have special property: The
definition of the Casimir energy for these black holes is similar
to that of the Kerr-Newman black hole, but the Cardy-Verlinde
formula takes the same form as that of the Kerr black hole.
Furthermore, we also study the entropy bounds for the systems in
which the matters surrounds these black holes. We find that the
bound for the case of the Kerr-Newman black hole is related to
its charge, and the bound for the cases of the EMDA, Kaluza-Klein,
and Sen black holes can be expressed as a unified form. A
surprising result is that the entropy bounds  for the
Kaluza-Klein and Sen black holes are tighter than the Bekenstein
one.

\end{abstract}

\vspace*{1.4cm}
 \pacs{ PACS numbers: 04.70.Dy, 04.20.-q, 97.60.Lf.}

\section{INTRODUCTION}
\label{sec:intro} \vspace*{0.2cm}

In a recent paper \cite{Verlinde00} Verlinde made an interesting
proposal that the $(1+1)$-dimensional Cardy formula \cite{Cardy}
can be generalized to the case in arbitrary dimensions. He argued
that the entropy of the conformal field theory (CFT) in a
spacetime $
 ds^2=-dt^2+R^2d\Omega^2_n,
 $
can be explicitly expressed as \cite{Verlinde00}
 \ba
 S=\frac{2\pi R}{\sqrt{a_1 b_1}}\sqrt{E_c(2E-E_c)}, \label{scv}
 \ea
where $E$ is the total energy, $E_c$ is the Casimir energy, and
$a_1$ and $b_1$ are two positive coefficients which are
independent of $R$ and $S$. For strong coupled CFT's with AdS
dual, the value of $\sqrt{a_1 b_1}$ is fixed to $n$ exactly. The
expression (\ref{scv}) is now referred to as the Cardy-Verlinde
formula.

The study of validity of the Cardy-Verlinde formula (\ref{scv})
for every typical spacetimes has attracted much attention recently
\cite{Verlinde00}, \cite{Cai01} -\cite{Cai01b} since it has not
been proved for all CFT's exactly yet: For the AdS Schwarzschild
black holes in various dimensions the formula (\ref{scv}) holds
exactly \cite{Verlinde00}; The Cardy-Verlinde formula is valid
for the  AdS Reissner-Nordstr\"om black hole if we subtract the
electric potential energy from the Casimir and total energies
\cite{Cai01}, but it is invalid for the AdS black holes in higher
derivative gravity  \cite{Cai01};  The formula holds for the
Taub-Bolt-AdS spacetimes at high temperature limit
\cite{Birmingham}; Klemm, Petkou, and Siopsis \cite{Klemm01}
showed that the Cardy-Verlinde formula holds for the Kerr-AdS$_n$
black holes; Motivating by the observational evidence that our
universe has positive cosmological constant
\cite{Perlmutter,Caldwell,Garnavich} and an interesting proposal
for dS/CFT correspondence \cite{Strominger}, Danielsson
\cite{Danielsson01}, Cai \cite{Cai01a}, Medved
\cite{Medved01,Medved01a}, and Ogushi \cite{Ogushi01} {\it et al}
attempt to generalize the formula to the case of the de Sitter
(dS) black holes; We \cite{Jing02} verified the Cardy-Verlinde
formula by using the Kerr-Newman-AdS$_4$ and Kerr-Newman-dS$_4$
black holes and found that it holds for these black holes if we
modify the definitions of the Casimir energy and the extensive
energy.

An interesting question, whether the Cardy-Verlinde formula holds
for a more general setting, e. g. for the black holes that are
asymptotically flat rather than approaching AdS/dS space, was
proposed by Klemm, Petkou, and Zanon \cite{Klemm011}. They argued
that for the asymptotically flat Schwarzschild and Kerr black
holes in any dimensions the Cardy-Verlinde formula can be
expressed as
 \ba
 S=\frac{2\pi r_+}{n}\sqrt{E_c\cdot 2E}, \label{kcv}
 \ea
where $r_+$ is the radius of the event horizon. However, at the
moment the question whether or not the Cardy-Verlinde formula can
be generalized to the asymptotically flat rotating charged black
holes obtained from the Einstein-Maxwell theory and low-energy
effective field theory describing string still remains open. The
main aim of this paper is to study the question by using some
typical spacetimes, such as the Kerr-Newman,
Einstein-Maxwell-dilaton-axion (EMDA), Kaluza-Klein, and Sen
black holes. The another purpose of the paper is to study the
entropy bounds for the systems in which the matters surround
these black holes.

The paper is organized as follows. In Sec. II, we study the
Cardy-Verlinde formula and entropy bound for the Kerr-Newman black
hole. In Sec. III, we consider the same question for the case of
the EMDA black hole. In Sec. IV, we  investigate  the case of the
Kaluza-Klein black hole. In Sec. V,  we discuss the case of the
Sen black hole. We present some discussions and conclusions in the
last section.

\vspace*{0.4cm}
\section{The Kerr-Newman black hole}
\vspace*{0.5cm}

The general charged stationary axisymmetric black hole obtained
from the Einstein-Maxwell gravitational theory is the Kerr-Newman
black hole. Mazur \cite{Mazur82} showed that the Kerr-Newman
solution is the only stationary, axisymmetric electrovac solution
of the Einstein-Maxwell equations. In the Boyer-Lindquist
coordinates the metric of the black hole reads \cite{Newman65}
\begin{eqnarray}
ds^{2} &=&-\left(\frac{\Delta-a^2 sin ^{2}\theta}{\rho^2}
\right)dt^2-\frac{2 a \sin^2\theta(r^2+a^2-\Delta)}{\rho^2}dt
d\phi
\nonumber  \\
&&+ \frac{\rho ^{2}}{\Delta} dr^{2}+\rho ^{2} d\theta
^{2}+\left[\frac{(r^2+a^2)^2-\Delta a^2
\sin^2\theta}{\rho^2}\right]\sin^2\theta d\phi^2, \label{kadsmet}
\end{eqnarray}
with
\begin{eqnarray}
\Delta &=&r^{2}+a^{2}-2Mr+Q^2,  \nonumber \\
\rho ^{2} &=&r^{2}+a^{2}\cos ^{2}\theta,  \label{delt}
\end{eqnarray}
where the parameters $M$, $a$, and $Q$ represent the mass, angular
momentum, and electric charge of the black hole, respectively.
The event horizon is located at $r_+$, which is determined by the
largest root of the $ \Delta _{r}=0 $. The mass $M $ can be
expressed as
\begin{equation}
M=\frac{r_+^2+a^2+Q^2}{2r_+}.
\end{equation}
Euclideanizing the metric (\ref{kadsmet}) and identifying
$\tau\sim \tau+ \beta $ and $\phi\sim \phi+i\beta \Omega_H$, we
can get the inverse Hawking temperature and the angular velocity
of the event horizon
 \ba
 \beta &=&\frac{4\pi r_+
(r_+^2+a^2)}{r_+^2-a^2-Q^2}, \nonumber \\
 \Omega_H&=&\frac{a}{r_+^2+a^2}. \label{omg}
 \ea
The angular momentum, electric potential and Bekenstein-Hawking
entropy are
 \ba
 {\cal{J}}&=&M a, \nonumber \\
 \Phi_Q&=&\frac{Q r_+}{r_+^2+a^2}, \nonumber \\
S&=&\pi (r_+^2+a^2). \label{s1}
 \ea
An electric potential when the rotation parameter $a$ goes to be
zero is given by
 \ba
 \Phi_{Q0}=\lim_{a\rightarrow 0}\Phi_Q=\frac{Q}{r_+}.\label{phi0}
 \ea

For the rotating black hole we define the Casimir energy as
 \ba
 E_c=n\left(E+pV-TS-{\mathcal{J}}\Omega_H-\frac{Q\Phi_Q}{2}
 -\frac{Q \Phi_{Q0}}{2}\right),\label{ecas}
 \ea
where the pressure is defined as $ p=-\left( \frac{\partial E }
{\partial V } \right)_{S,J,Q} $ and $n=2$. The definition of the
Casimir energy (\ref{ecas}) takes the same form as that of the
Kerr-Newman-AdS$_4$ and Kerr-Newman-dS$_4$ black holes
\cite{Jing02}. Substituting Eqs. (\ref{omg}), (\ref{s1}), and
(\ref{phi0}) into Eq. (\ref{ecas}), we get
 \ba
 E_c=\frac{(r_+^2+a^2) }{r_+}. \label{ecas01}
 \ea
We also define the extensive energy as
 \ba
E_{ext}=2\left(E-\frac{Q \Phi_{Q0}}{2}\right)=2(E-E_Q),
\label{eex}
 \ea
where $E_Q=\frac{Q\Phi_{Q0}}{2}$. In this case the extensive
energy is
 \ba
 E_{ext}=\frac{(r_+^2+a^2)}{r_+}. \label{eex01}
 \ea
We should note that the definition of the extensive energy
(\ref{eex}) is different from that of the Kerr-Newman-AdS$_4$ and
Kerr-Newman-dS$_4$ black holes \cite{Jing02} by a term $E_c$.
With the Casimir energy (\ref{ecas01}) and the extensive energy
(\ref{eex01}), we find that the entropy of the CFT on the horizon
can be written as
 \ba
S&=&\frac{2\pi r_+}{n}\sqrt{E_c\left[2 \left(E-E_Q\right)\right]
}\nonumber \\
&=&\pi (r_+^2+a^2),  \label{cv}
 \ea
which is equal to the Bekenstein-Hawking entropy of the
Kerr-Newman black hole. The result in this section reduces to that
of the Kerr black hole obtained in Ref. \cite{Klemm011} when the
charge $Q$ becomes zero.

We now study the entropy bound for a system in which some matters
surround the Kerr-Newman black hole. We can define the entropy
$S_B$ which relates the energy $E-E_Q$ by
 \ba
 S_B=2\pi (E-E_Q) R, \label{knsb}
 \ea
where $R$ is the radius of the sphere circumscribing the system
($R\geq r_+$). Substituting Eq. (\ref{eex}) into (\ref{knsb}) we
get
 \ba
 S_B=\pi (r_+^2+a^2)\frac{R}{r_+}=S\frac{R}{r_+},
 \ea
which shows that
 \ba
 S\leq S_B=2\pi R\left(E-\frac{Q^2}{2r_+}\right).
 \ea
It reduces to the Bekenstein bound when the charge $Q$ becomes
zero.

\vspace*{0.4cm}
\section{The Stationary axisymmetric EMDA black hole}
\vspace*{0.5cm}

The general four-dimensional low-energy action obtained from
string theory is \cite{Garcia}
  \ba
 \label{emaction}
     I&=&\frac{1}{16\pi}\int{\rm d}^{4}x \sqrt{-{\rm g}}
     ({\cal{R}}-2{\rm g}^{\mu\nu}
         \bigtriangledown_\mu\Phi\bigtriangledown_\nu\Phi
         -\frac{1}{2}{\rm e}^{4\Phi}{\rm g}^{\mu\nu}
         \bigtriangledown_\mu{\rm K}_{\rm a}\bigtriangledown_\nu
         {\rm K}_{a} \nonumber \\
      & &-{\rm e}^{-2\Phi}{\rm g}^{\mu\lambda}
         {\rm g}^{\nu\rho}{\rm F}_{\mu\nu}{\rm F}_{\lambda\rho}
        -{\rm K}_{a}{\rm F}_{\mu\nu}
         \tilde{{\rm F}}^{\mu\nu}),
  \ea
with $ \tilde {F}_{\mu\nu}=-\frac{1}{2} \sqrt{-{\rm g}}
\varepsilon_{\mu\nu\alpha\beta} F^{\alpha\beta} $,  where ${\cal{
R}}$ is the scalar Riemann curvature, $\Phi$ is the massless
dilaton field, $ F_{\mu\nu}$ is the electromagnetic antisymmetric
tensor field, and $K_{\rm a}$ is the axion field dual to the
three-index antisymmetric tensor field  $ H=-\exp(4\Phi)*{\rm
d}K_{\rm a}/4 $.

The stationary axisymmetric EMDA black hole solution get from the
action (we take the solution $ b=0$ in Eq.(14) in
Ref.\cite{Garcia}; the reason we use this solution is that the
solution $b\not=0$ cannot be interpreted properly as a black
hole) is given \cite{Garcia} by
 \ba \label{emdamet}
 {\rm d}s^2&=&-\frac{\Delta-a^2\sin^2\theta}{\rho^2}
 {\rm d}t^2-\frac{2a\sin^2\theta}{\rho^2}\left[
   (r^2+a^2-2Dr)-\rho^2 \right ]
   {\rm d}t{\rm d}\phi\nonumber \\
 & &+\frac{\rho^2}{\Delta}{\rm d}r^2+\rho^2 {\rm d}{\theta}^2
   +\frac{\sin^2\theta}{\rho^2}\left [(r^2+a^2-2Dr)^2-\Delta a^2
   \sin^2\theta\right ]
   {\rm d}\phi^{2},
 \ea
 with
 \ba  \Delta=r^2-2mr+a^2,\;\;\;\;\;
\rho^2=r^2-2Dr+a^2\cos^2\theta,
 \ea
 and
 \ba {\rm
e}^{2\Phi}&=&\frac{W}{\Delta}=\frac{\omega}{\Delta}(r^2+a^2\cos^2\theta),
\;\;\; \omega={\rm e}^{2\Phi_{0}}, \nonumber \\
K_a&=&K_0+\frac{2aD\cos\theta}{W}, ~~~~~~ A_t=\frac{1}{\Delta}(Q
r-{\rm g}a\cos\theta),~~~~~~A_r=A_\theta=0,\nonumber \\
A_{\phi}&=&\frac{1}{a\Delta}(-Qra^2\sin^2\theta+{\rm
g}(r^2+a^2)a\cos\theta).
 \ea
The mass $M$, angular momentum ${\cal{J}}$, electric charge $Q$,
and magnetic charge $P$ of the black hole are given by
 \ba \label{mjq}
M&=&m-D,~~~~~~ {\cal{J}}=a(m-D),\nonumber \\
Q&=&\sqrt{2\omega D(D-m)},~~~~~~P={\rm g}.
 \ea
The inverse Hawking temperature and the angular velocity are
 \ba
 \beta &=&\frac{4\pi r_+
(r_+^2+a^2-2Dr_+)}{(r_+^2-a^2)}, \nonumber \\
 \Omega_H&=&\frac{a}{r_+^2+a^2-2Dr_+}. \label{emdaomg}
 \ea
The electric potential and the Bekenstein-Hawking entropy of the
black hole are
 \ba
 \Phi_Q&=&\frac{Qr_+}{\omega (r_+^2+a^2-2Dr_+)}, \nonumber \\
 S&=&\pi (r_+^2+a^2-2Dr_+). \label{emdas1}
 \ea

We now define the Casimir energy as
 \ba
 E_c=n\left(E+pV-TS-{\mathcal{J}}\Omega_H-\frac{Q\Phi_Q}{2}
 \right),\label{emdaecas}
 \ea
where $n=2$ and  the pressure is defined as $ p=-\left(
\frac{\partial E } {\partial V } \right)_{S,J,Q} $. Substituting
quantities of the EMDA black hole into Eq. (\ref{emdaecas}), we
arrive at
 \ba
 E_c=\frac{(r_+^2+a^2-2Dr_+) }{r_+}=2E. \label{emdaecas01}
 \ea
Thus the entropy of the CFT can be cast as
 \ba
S&=&\frac{2\pi r_+}{n}\sqrt{E_c\cdot 2E}\nonumber \\
&=&\pi (r_+^2+a^2-2Dr_+). \label{emdacv}
 \ea
It is of interest to note that the definitions of the Casimir
energy (\ref{emdaecas}) of the EMDA black hole has similar form as
that of the Kerr-Newman black hole, but the Cardy-Verlinde
formula (\ref{emdacv}) takes the same form as that of the Kerr
black hole (\ref{kcv}). We should not be surprised at this fact
because the charged EMDA black hole possesses several different
properties as compared to the charged Kerr-Newman black hole: (a)
The horizons of the Kerr-Newman black hole are given by
$r_{\pm}=M\pm\sqrt{M^2-a^2-Q^2}$, whereas in the case of the EMDA
black hole we have $r_{\pm}=m \pm\sqrt{m ^2-a^2}$ which takes the
same form as that of the Kerr black hole; (b) The heat capacity
of the Kerr-Newman black hole is $C_{QJ}=\frac{MS^3T}{\pi
{\cal{J}}^2+\pi Q^2/4-S^3T^2}$, while for the EMDA black hole we
get $C_{QJ}=\frac{MS^3T}{\pi {\cal{J}}^2-S^3T^2}$ \cite{Jing96}
which also has the same form as that of the Kerr black hole.

We now study the entropy bound for a system in which some matters
surround the black hole. We can define an entropy $S_B$ which
relates the energy $E$ by
 \ba
 S_B=2\pi E R=\pi (r_+^2+a^2-2Dr_+)\frac{R}{r_+}
 =S\frac{R}{r_+}. \label{EMDAsb}
 \ea
Thus we have
 \ba
 S\leq S_B=2\pi R E.
 \ea
Which takes the same form  as that of  the Bekenstein bound
(\ref{kcv}).

\vspace*{0.4cm}
\section{The stationary Kaluza-Klein black hole}
\vspace*{0.5cm}

 The four dimensional low-energy action obtained
from string theory is \cite{Frolov87}
\begin{equation}\label{action1}
I=\frac{1}{2 \kappa }\int d^4x\sqrt{-g}\left[{\cal R}-2(\nabla
\phi)^2-e^{-2\alpha\phi}F^{2}\right],
\end{equation}
where  $\phi$  is the dilaton scalar field, $F_{ab}$ is the
Maxwell field, and $\alpha$ is a free parameter which governs the
strength of the coupling of the dilaton to the Maxwell field. The
stationary Kaluza-Klein black hole is described by a solution of
the motion equations obtained from Eq. (\ref{action1})  with
$\alpha=\sqrt{3}$, which takes the form \cite{Frolov87}
 \ba
 ds^2=&-&\frac{1-Z}{B}d t^2-\frac{2aZ \sin^2\theta}{B\sqrt{1-v^2}}dt
 d\phi+\frac{B\rho^2}{\triangle_r}
 d r^2+B \rho^2 d \theta ^2\nonumber \\
 &+&\left[B(r^2+a^2)+a^2\sin^2\theta
 \frac{Z}{B}\right] \sin^2\theta d\phi^2,\label{kkmetric}
 \ea
with
 \ba
 Z&=&\frac{2mr}{\rho^2},\ \ \ B=\left(1+\frac{v^2
 Z}{1-v^2}\right)^{1/2}, \nonumber \\
 \rho^2&=&r^2+a^2\cos^2\theta, \ \ \triangle_r=r^2+a^2-2mr,
 \ea
where $a$ and $v$ are the rotation parameter and the velocity of
the boost. The dilaton scalar field, and the vector potential of
the stationary Kaluza-Klein black hole are
\begin{eqnarray}
\Phi=-\frac{\sqrt{3}}{2}\ln B,  \ \
A_t=\frac{v}{2(1-v^2)}\frac{Z}{B^2}, \ \ A_\phi=-\frac{a v
\sin^2\theta}{2\sqrt{1-v^2}}\frac{Z}{B^2}. \label{kl}
\end{eqnarray}
We get the inverse Hawking temperature and the angular velocity
 \ba
 \beta &=&\frac{4\pi r_+
(r_+^2+a^2)}{(r_+^2-a^2)\sqrt{1-v^2}}, \nonumber \\
 \Omega_H&=&\frac{a\sqrt{1-v^2}}{r_+^2+a^2}. \label{kkomg}
 \ea
The ADM mass, angular momentum ${\cal{J}}$, charge $Q$, and
corresponding potential of the black hole are given by
 \ba
M&=&\left[1+\frac{v^2}{2(1-v^2)}\right]m, \nonumber \\
{\cal{J}}&=&\frac{m a}{\sqrt{1-v^2}}, \nonumber \\
 Q&=&\frac{m v}{1-v^2}, \nonumber \\
 \Phi_Q&=&
 \frac{Qr_+(1-v^2)}{r_+^2+a^2},\label{kkam}
 \ea
while the Bekenstein-Hawking entropy of the black hole is
 \ba
S=\frac{\pi (r_+^2+a^2)}{\sqrt{1-v^2}}. \label{kks1}
 \ea

As the EMDA black hole we define the Casimir energy as
 \ba
 E_c=n\left(E+pV-TS-{\mathcal{J}}\Omega_H-\frac{Q\Phi_Q}{2}
\right),\label{kkecas}
 \ea
where $n=2$ and the pressure is defined as $ p=-\left(
\frac{\partial E } {\partial V } \right)_{S,J,Q} $. Substituting
Eqs. (\ref{kkomg}), and (\ref{kks1}) into Eq. (\ref{kkecas}), we
get
 \ba
 E_c=\frac{(r_+^2+a^2)(2-v^2) }{2 r_+ (1-v^2)}. \label{kkecas01}
 \ea
The extensive energy is defined as
 \ba
E_{ext}=2 E=\frac{(r_+^2+a^2)(2-v^2)}{2 r_+ (1-v^2)}.
\label{kkeex}
 \ea
Then, the entropy of the CFT can be cast into
 \ba
S&=&\frac{2\pi r_+}{\sqrt{a_1 b_1}}\sqrt{E_c\cdot 2E
}\nonumber \\
&=&\frac{\pi (r_+^2+a^2)}{\sqrt{1-v^2}}, \label{kkcv}
 \ea
where
 \ba
a_1=b_1=\frac{n\left(1-\frac{v^2}{2}\right)}
{\sqrt{1-v^2}}.\label{cons}
 \ea
The definition of the Casimir energy (\ref{kkecas}) and the
Cardy-Verlinde formula (\ref{kkcv}) possess the same form as that
of the EMDA black hole. The entropy of the CFT (\ref{kkcv})
agrees exactly with the Bekenstein-Hawking entropy (\ref{kks1}).

\begin{figure}[t]
\centerline{
        \psfig{figure=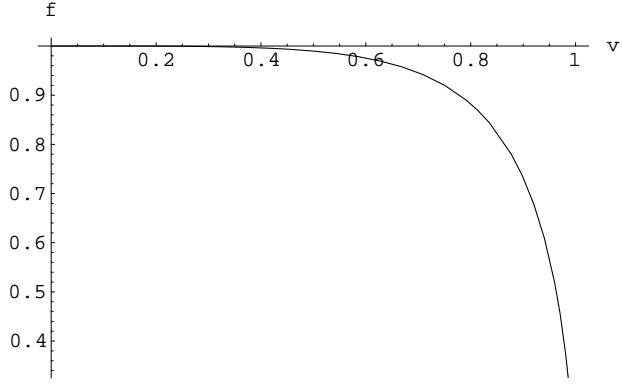,height=2.0in,angle=0}}
        \vspace*{1.0cm}
        \caption{
        In the figure $f=\frac{2}{\sqrt{a_1 b_1}}
        =\frac{\sqrt{1-v^2}}{1-v^2/2}$.
        }
\label{fig1}

\end{figure}

To study the entropy bound for a system in which some matters
surround the Kaluza-Klein black hole, we can define an entropy
$S_B$ as
 \ba
 S_B=2\pi E R \left( \frac{n}{\sqrt{a_1 b_1}}\right)
 =\frac{\pi (r_+^2+a^2)}{\sqrt{1-v^2}}\frac{R}{r_+}
 =S\frac{R}{r_+}. \label{kksb}
 \ea
Noting that $R\geq r_+$ we obtain
 \ba
 S\leq S_B=2\pi R E \left(\frac{n}{\sqrt{a_1 b_1}}\right).
 \label{kksb1}
 \ea
The entropy bound is different from the Bekenstein bound by a
factor $ \frac{n}{\sqrt{a_1 b_1}}$. We know from the Fig.
(\ref{fig1}) that
 \ba
 \frac{n}{\sqrt{a_1 b_1}} \leq 1.
 \ea
The relation shows that the entropy bound (\ref{kksb1}) is
tighter than the Bekenstein bound $S\leq 2\pi E R $
\cite{Bekenstein}.
 \vspace*{1.0cm}

\vspace*{0.4cm}
\section{The Sen black hole in
heterotic string}
\vspace*{0.5cm}

The  effective action of a toroidally compactified heterotic
string in $D$-dimensions takes the form\cite{Maharana93,Sen94}
 \ba
I&=&\frac{1}{2 \kappa }\int d^Dx \sqrt{-g}\left[{\cal R}-{1\over
(D-2)} g^{\mu\nu}\partial_{\mu}\Phi\partial_{\nu}\Phi+{1\over 8}
g^{\mu\nu}{\rm Tr}(\partial_{\mu}ML\partial_{\nu}ML)\right.\cr &
&-\left.{1\over{12}}
e^{-2\alpha\phi}g^{\mu\mu^{\prime}}g^{\nu\nu^{\prime}}
g^{\rho\rho^{\prime}}H_{\mu\nu\rho}H_{\mu^{\prime}\nu^{\prime}
\rho^{\prime}}  -{1\over
4}e^{-\alpha\Phi}g^{\mu\mu^{\prime}}g^{\nu\nu^{\prime}} {\cal
F}^{i}_{\mu\nu}(LML)_{ij} {\cal
F}^{j}_{\mu^{\prime}\nu^{\prime}}\right], \label{effaction} \ea
 where
$g\equiv {\rm det}\,g_{\mu\nu}$, ${\cal R}_g$ is the Ricci scalar
of $g_{\mu\nu}$, $\Phi$ is the dilaton field, ${\cal F}^i_{\mu\nu}
=
\partial_{\mu} {\cal A}^i_{\nu}-\partial_{\nu} {\cal A}^i_{\mu}$
are the $U(1)^{36-2D}$ gauge field strengths,
$H_{\mu\nu\rho}=(\partial_\mu B_{\nu\rho}+2{\cal
A}_{\mu}^{i}L_{ij}{\cal F}^{j}_{\nu\rho})+{\rm cyclic \
permutations\ of} \ \mu, \nu, \rho,$
 and $M$ is the O(10-D,
26-D) symmetric matrix and $L$ is
\begin{equation} L =\left ( \matrix{0 & I_{10-D}& 0\cr I_{10-D}
& 0& 0 \cr 0 & 0 &  I_{26-D}} \right ). \label{4dL}
\end{equation}
The metric of the Sen black hole is \cite{Sen95}
 \ba \label{e34}
ds^2 &=& \Delta^{1\over 2} \Big\{ - \Delta^{-1}
(r^2+a^2\cos^2\theta -2m r) dt^2 + (r^2+a^2-2m r)^{-1} d r^2
+d\theta^2 \nonumber \\
&& +\Delta^{-1} \sin^2\theta [\Delta + a^2\sin^2\theta (r^2 +
a^2\cos^2 \theta + 2m r \cosh\alpha\cosh\beta)] \, d\phi^2
\nonumber \\  && - 2\Delta^{-1} m r a \sin^2\theta (\cosh\alpha +
\cosh\beta) dt d\phi\Big\}\, , \ea
 with
 \be \label{e27} \Delta = (r^2 + a^2\cos^2\theta)^2 + 2m
r (r^2 + a^2 \cos^2\theta) (\cosh\alpha \cosh\beta -1) + m^2 r^2
(\cosh\alpha -\cosh\beta)^2\, , \ee where $\alpha$ and $\beta$
are two boosts, and $a$ and $m$ represent the rotational and mass
parameters, respectively. The black hole will be called Sen black
hole for short.  The two horizons of the black hole are shown by
  \be
\label{e40} r_+ = m \pm \sqrt{m^2 - a^2}  .
 \ee
The surface gravity of the black hole is given by, \be
\label{e42} \kappa = { \sqrt{m^2 - a^2} \over m(\cosh\alpha +
\cosh\beta) (m+\sqrt{m^2 - a^2}) }\, .
  \ee
$\kappa/2\pi$ can be interpreted as the Hawking temperature of
the black hole.  The angular velocity $\Omega$ at the horizon is
 \be \label{e44}
\Omega _H = {a\over m (\cosh\alpha + \cosh\beta) (m + \sqrt{m^2 -
a^2})}\, . \ee
 The mass $M$, angular momentum ${\cal{J}}$,
electric charge $Q^{a}$ and magnetic dipole moment $\mu^{a}$
given by
 \ba \label{e35} M &=& {1\over 2} m ( 1 +
\cosh\alpha \cosh \beta)\, , \nonumber \\
   {\cal{J}}& =& {1\over 2} ma
(\cosh\alpha + \cosh \beta) \, , \nonumber \\
  Q^{a} &=&
{m\over \sqrt 2} \sinh\alpha \cosh\beta \, n^a \qquad~~~~~~~~
\hbox{for}
\quad 1\le a \le 22\, \nonumber \\
&=& {m\over \sqrt 2} \sinh\beta \cosh\alpha \, p^{(a-22)} \qquad
~~\hbox{for} \quad 23\le a\le 28\, , \nonumber \\
  \mu^{a} &=&
{1\over \sqrt 2} ma \sinh\alpha \, n^a \qquad
~~~~~~~~~~~~\hbox{for}
\quad 1\le a \le 22\, \nonumber \\
&=& {1\over \sqrt 2} ma \sinh\beta \, p^{(a-22)} \qquad
~~~~~~~\hbox{for} \quad 23\le a\le 28\, .
 \ea
The Bekenstein-Hawking entropy of the black hole is
 \be \label{e41}
S = \pi m (\cosh\alpha +\cosh\beta) \, (m+\sqrt{m^2 - a^2})\, .
 \ee
By setting the boost parameter $\beta=0$, the metric (\ref{e34})
reduces to the black hole in Ref. \cite{Sen} and Eq. (\ref{e35})
becomes \cite{Sen}
  \ba
\label{9}
 M &=& \frac{m(1+\cosh\alpha)}{2},~~~~
 {\cal{J}}=\frac{ma(1+\cosh\alpha)}{2}, \nonumber\\
Q &=& \frac{m\sinh\alpha}{ \sqrt{2}}, ~~~~~~
\mu=\frac{ma\sinh\alpha}{ \sqrt{2}}.
  \ea
See Eq.(16) and Eq.(17) of Ref.\cite{Sen}. This is a special case
of the Sen black hole, so the discussion in this section will be
valid for the black hole.

The Casimir energy can also be define as
 \ba
 E_c=n\left(E+pV-TS-{\mathcal{J}}\Omega_H-\frac{Q\Phi_Q}{2}
 \right),\label{hsecas}
 \ea
where $n=2$ and  the pressure is defined as $ p=-\left(
\frac{\partial E } {\partial V } \right)_{S,J,Q} $. Substituting
quantities of the black hole into Eq. (\ref{hsecas}), we arrive at
 \ba
 E_c=\frac{(r_+^2+a^2)(1+\cosh \alpha
 \cosh\beta) }{2 r_+}=2E. \label{hsecas01}
 \ea
Therefore, the entropy of the CFT can be expressed as
 \ba
S&=&\frac{2\pi r_+}{\sqrt{a_1 b_1}}\sqrt{E_c\cdot 2E}\nonumber \\
&=&\frac{\pi}{2} (r_+^2+a^2)(\cosh{\alpha}+\cosh{\beta}),
\label{hscv}
 \ea
where
 \ba
 a_1=b_1=\frac{n(1+\cosh{\alpha}\cosh{\beta})}
 {\cosh{\alpha}+\cosh{\beta}}.
 \ea
The entropy of the CFT (\ref{hscv}) is equal to the
Bekenstein-Hawking entropy (\ref{e41}). The the Casimir energy
(\ref{hsecas}) and the Cardy-Verlinde formula (\ref{hscv}) for
the Sen black hole also take the same form as that of the EMDA
and Kaluza-Klein black holes. The result can be used for the black
hole in Ref. \cite{Sen}, i. e., for the case $\beta=0$.

\begin{figure}[t]
\centerline{
        \psfig{figure=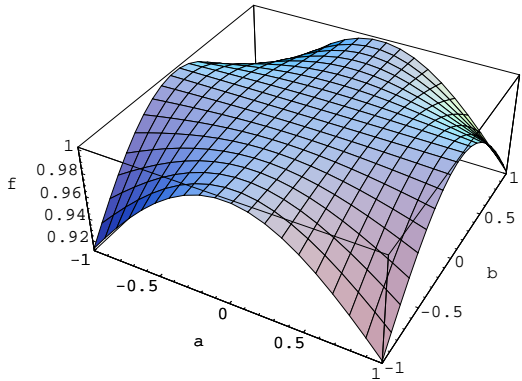,height=2.0in,angle=0}}
        \vspace*{1.0cm}
        \caption{
        In the figure $ f=\frac{2}{\sqrt{a_1 b_1}}=\frac{
        \cosh \alpha +\cosh \beta }{1+\cosh \alpha \cosh \beta}$,
        $a$ and $b$ represent $\alpha$ and $\beta$, respectively.
        }
\label{fig2}

\end{figure}

For a system in which matter surrounds the Sen black hole, we can
define the entropy $S_B$ which relates the energy $E$ by
 \ba
 S_B&=&2\pi E R \left( \frac{n}{\sqrt{a_1 b_1}}\right) \nonumber \\
 &=&\frac{\pi (r_+^2+a^2)(\cosh \alpha +\cosh \beta )}{2}
 \frac{R}{r_+}\nonumber \\
 &=&S\frac{R}{r_+}. \label{ssb}
 \ea
From which we obtain the entropy bound
 \ba
 S\leq S_B=2\pi R E \left(\frac{n}{\sqrt{a_1 b_1}}\right).\label{ssb1}
 \ea
The Fig. (\ref{fig2}) shows that
 \ba
 \frac{n}{\sqrt{a_1 b_1}} \leq 1.
 \ea
Therefore, as the case of the Kaluza-Klein black hole, we also
note that the entropy bound (\ref{ssb1}) for the Sen black hole is
tighter than the Bekenstein one.

\vspace*{1.0cm}

\vspace*{0.4cm}
\section{conclusion and discussion}
\vspace*{0.5cm}

The Cardy-Verlinde formula, which relates the entropy of the
conformal field theory to its Casimir energy $E_c$, the energy
$E$ (or the extensive energy), and the radius $R$ of the unit
sphere $S^n$, is generalized to the asymptotically flat rotating
charged black holes in the Einstein-Maxwell theory and low-energy
effective field theory describing string, such as the
Kerr-Newman, EMDA, Kaluza-Klein, and Sen black holes.

For the Kerr-Newman black hole, we find that the definition of the
Casimir energy (\ref{ecas}) takes the same form as that of the
Kerr-Newman-AdS$_4$ and Kerr-Newman-dS$_4$ black holes
\cite{Jing02}, while the extensive energy (\ref{eex}) is
different from that of the Kerr-Newman-AdS$_4$ and
Kerr-Newman-dS$_4$ black holes \cite{Jing02}. The Cardy-Verlinde
formula for the  black hole can be generalized as
 $
 S=\frac{2\pi r_+}{n}\sqrt{E_c\left[2
\left(E-E_Q\right)\right]}.
 $

The charged EMDA, Kaluza-Klein, and Sen black holes possess some
special properties as compare to the Kerr-Newman black hole. The
definition of the Casimir energies (\ref{emdaecas}),
(\ref{kkecas}), and (\ref{hsecas}) have the similar form as that
of the Kerr-Newman black hole, but the Cardy-Verlinde formulae
(\ref{emdacv}), (\ref{kkcv}), and (\ref{hscv})  take the same form
as that of the Kerr black hole \cite{Klemm011}, which can be
written as follows a united form $
 S=\frac{2\pi r_+}{\sqrt{a_1b_1}}\sqrt{E_c\cdot 2E},
 $
where $\sqrt{a_1 b_1}=n$ for the EMDA black hole, $\sqrt{a_1
b_1}=\frac{n\left(1-v^2/2\right)} {\sqrt{1-v^2}}$ for the
Kaluza-Klein black hole, and
 $\sqrt{a_1 b_1}=\frac{n(1+\cosh{\alpha}\cosh{\beta})}
 {\cosh{\alpha}+\cosh{\beta}}$ for the Sen black hole.
The result is also valid for the black hole in Ref. \cite{Sen}
since it can be considered as a special case of the Sen black
hole.

Another important conclusion obtained for these asymptotically
flat rotating charged black holes is that the entropies of the
CFTs agree precisely with their Bekenstein-Hawking entropies.

For stationary charged black holes in the Einstein-Maxwell theory
and the low-energy effective field theory describing string, by
using the covariant phase technique
\cite{Lee90,Wald93,Iyer94,Iyer95} and Carlip's boundary
\cite{Carlip99l,Carlip99} conditions, we constructed
\cite{Jing,Jing01a} the standard Virasoro subalgebra with
corresponding central charge at the Killing horizons and proved
that the density of states determined by conformal fields theory
methods yields the statistical entropy which agrees with the
Bekenstein-Hawking entropy. Comparing the result with that
obtained in this paper, it is surprising to note that different
ways arrive at the same conclusion!

Furthermore, we also study the entropy bounds for the systems in
which the matters surrounds these black holes. For the case of the
Kerr-Newman black hole, the bound $ S\leq 2\pi
R\left(E-\frac{Q^2}{2r_+}\right)$ is tightened by the electric
charge. The bound reduces to the Bekenstein bound when the charge
$Q$ becomes zero. For the case of the charged EMDA, Kaluza-Klein,
and Sen black holes, the entropy bound can be expressed as a
unified form, $S\leq 2\pi E R \frac{n}{\sqrt{a_1 b_1}}$. It is of
interest to note that the entropy bound  for the Kaluza-Klein and
Sen black holes is tighter than the Bekenstein one since
$\frac{n}{\sqrt{a_1 b_1}}\leq 1$ for these cases.

\vspace*{0.8cm}

\begin{acknowledgements}
This work was supported by the National Natural Science Foundation
of China under Grant No. 19975018, and Theoretical Physics Special
Foundation of China under Grant No. 19947004.
\end{acknowledgements}

\end{document}